\documentclass{article}
\usepackage{amsfonts}
\usepackage{amsmath}
\usepackage{cite}

\input{tcilatex}
\begin{document}

\title{Complete non-relativistic bound state solutions of the Tietz-Wei
potential via the path integral approach}
\author{A. Khodja, A. Kadja, F. Benamira and L. Guechi \\
%EndAName
Laboratoire de Physique Th\'{e}orique, D\'{e}partement de Physique, \and %
Facult\'{e} des Sciences Exactes, Universit\'{e} des fr\`{e}res Mentouri,
\and Route d'Ain El Bey, Constantine, Algeria}
\maketitle

\begin{abstract}
In this work, the bound state problem of some diatomic molecules in the
Tietz-Wei potential with varying shapes is correctly solved by means of path
integrals. Explicit path integration leads to the radial Green's function in
closed form for three different shapes of this potential. In each case, the
energy equation and the wave functions are obtained from the poles of the
radial Green's function and their residues, respectively. Our results prove
the importance the optimization parameter $c_{h}$ in the study of this
potential which has been completely ignored by the authors of the papers
cited below. In the limit $c_{h}\rightarrow 0$, the energy spectrum and the
corresponding wave functions for the radial Morse potential are recovered.


03.65.Db-Functional analytical methods


Corresponding author, e-mail: guechilarbi@yahoo.fr
\end{abstract}

\section{Introduction\qquad \qquad \qquad \qquad \qquad \qquad \qquad \qquad
\qquad \qquad \qquad \qquad \qquad \qquad \qquad \qquad \qquad \qquad \qquad
\qquad \qquad \qquad \qquad \qquad \qquad \qquad \qquad \qquad \qquad \qquad}

The four-parameter potential function so-called Tietz-Wei potential \cite%
{Tietz,Wei,Natanson} introduced to describe the bond-stretching vibrations
of diatomic molecules is defined by

\begin{equation}
V_{TW}(r)=D\left[ \frac{1-e^{-b_{h}(r-r_{e})}}{1-c_{h}e^{-b_{h}(r-r_{e})}}%
\right] ^{2};\text{ }b_{h}=\beta (1-c_{h}),  \label{3.1}
\end{equation}%
where $r$ is the internuclear distance, $D$ and $r_{e}$ represent the depth
of the potential well and the length of the molecular bond, respectively.
For physical reasons, the optimization parameter $c_{h}$ is a dimensionless
constant chosen such that $\left\vert c_{h}\right\vert <1$ and $\beta $ is
the Morse constant . This potential is much more realistic than the Morse
potential in the description of molecular dynamics. Since 1990, it continues
to interest physicists and chemists if we judge by the impressive number of
works done on its applications in molecular physics and quantum chemistry.
Using the Hamilton-Jacobi theory and the Bohr-Sommerfeld quantization rule,
the rotation-vibration energy levels of diatomic molecules \cite{Kunc} as
well as radial probability distributions of some diatomic molecules in
excited rotation-vibration states \cite{Gordillo} were obtained. In a
three-dimensional space, the analytical solution of the radial Schr\"{o}%
dinger equation with this potential was recently given for $s(l=0)-$waves in
a formulation based on the Nikiforov-Uvarov method \cite{Hamzavi1} and for
the $l-$waves using the approximation of Pekeris to replace the centrifugal
potential term \cite{Hamzavi2}. The approximate analytical solutions of the
Schr\"{o}dinger, Klein-Gordon and Dirac equations were also proposed by
adopting the Pekeris approximation \cite{Pekeris} and by applying three
types of techniques namely the asymptotic iteration method, the functional
analysis approach and the supersymmetry in quantum mechanics \cite{Falaye1}.
In another study, the construction of Ladder operators by the algebraic
approach and coherent states in the context of supersymmetry in quantum
mechanics for the Tietz-Wei anharmonic potential was discussed \cite%
{Mikulski}. The method based on the appropriate quantization rule has
recently been applied to determine the energy spectrum of a set of diatomic
molecules in the presence of the Tietz-Wei potential \cite{Falaye2} and the
modified Tietz-Wei potential \cite{Sameer2}. The Schr\"{o}dinger equation
has been solved by adopting the Pekeris approximation for the centrifugal
term. Finally, we find in a very recent work done in a space with $D$
dimensions the solutions of the Schr\"{o}dinger equation with the Tietz-Wei
potential obtained by using a Pekeris approximation and supersymmetry in
quantum mechanics \cite{Zarrinkamar}.

As in the articles quoted above, the parameter $c_{h}$ does not seem to have
any influence on the solutions of the problem, we will devote the rest of
this paper to a rigorous discussion by the path integral approach
considering all the existing cases and prove the importance of $c_{h}$ to
solve the problem properly.

In the second section, we express the radial Green's function with respect
to the potential $V_{TW}(r)$ in the form of a path integral, emphasizing
that the problem of its construction is presented differently according to
the value that takes the parameter $c_{h}$. We establish the correspondence
between the Tietz-Wei potential and the deformed Manning-Rosen potential
when $0<c_{h}<1$ in the third section. For $e^{-b_{h}r_{e}}\leq c_{h}<1$, by
replacing the centrifugal potential term with an appropriate approximation,
we compute, in a closed form, the radial Green's function associated with
the $l-$waves. From this we deduce the energy spectrum and the appropriately
normalized wave functions of the bound states by indicating the minimum
values {}{}of the parameter $c_{h}$ from which the numerical values {}{}of
the energy spectrum for certain diatomic molecules can be obtained. When $%
0<c_{h}<e^{-b_{h}r_{e}}$, we limit ourselves to the study of $s(l=0)-$waves
and we calculate the Green's function through the technique used to solve
problems with Dirichlet boundary conditions described in our previous work 
\cite{Khodja2} that it is not necessary to repeat here. In the fourth
section, we consider the case where $-1<c_{h}$ $<0$. The potential (\ref{3.1}%
) is reduced to the deformed Rosen-Morse potential. We evaluate the Green's
function associated with the $s(l=0)-$waves in the same way as before. In
these last two cases, we explicitly obtain the wave functions and we find a
transcendental equation to determine the energy spectrum. At the limit where 
$c_{h}$ $\rightarrow 0$, we recover the Green's function, the energy
spectrum and the wave functions for the Morse potential in the fifth
section. The sixth section will be a conclusion.

\section{Green's function}

Since the potential (\ref{3.1}) has spherical symmetry, the radial Green's
function is defined by%
\begin{equation}
G_{l}\left( r^{\prime \prime },r^{\prime };E\right) =\int_{0}^{\infty
}dT\left\langle r^{\prime \prime }\right\vert \exp \left\{ \frac{i}{\hbar }%
\left[ E-H_{l}\right] T\right\} \left\vert r^{\prime }\right\rangle ,
\label{3.2}
\end{equation}%
where $H_{l}$ is the Hamiltonian%
\begin{equation}
H_{l}=\frac{P_{r}^{2}}{2\mu }+V_{eff}(r),  \label{3.3}
\end{equation}%
with the effective potential%
\begin{equation}
V_{eff}(r)=\frac{\hbar ^{2}l(l+1)}{2\mu r^{2}}+V_{TW}(r).  \label{3.4}
\end{equation}

In the context of Feynman's path integral approach \cite{Peak}, it is easy
to express (\ref{3.2}) as a path integral,

\begin{equation}
G_{l}\left( r^{\prime \prime },r^{\prime };E\right) =\int_{0}^{\infty
}dT\exp \left( \frac{i}{\hbar }ET\right) K_{l}\left( r^{\prime \prime
},r^{\prime };T\right) ,  \label{3.5}
\end{equation}%
with 
\begin{eqnarray}
K_{l}\left( r^{\prime \prime },r^{\prime };T\right) &=&\lim_{N\rightarrow
\infty }\dprod_{n=1}^{N}\left[ \int dr_{n}\right] \dprod_{n=1}^{N+1}\left[
\int \frac{d\left( P_{r}\right) _{n}}{2\pi \hbar }\right]  \notag \\
&&\times \exp \left\{ \frac{i}{\hbar }\underset{n=1}{\overset{N+1}{\sum }}%
\left[ \left( P_{r}\right) _{n}\Delta r_{n}-\varepsilon \left( \frac{\left(
P_{r}\right) _{n}^{2}}{2\mu }+V_{eff}(r_{n})\right) \right] \right\} . 
\notag \\
&&  \label{3.6}
\end{eqnarray}%
Then, by integrating on the variables $\left( P_{r}\right) _{n}$, we obtain
the path integral in\ the configuration space,%
\begin{equation}
G_{l}\left( r^{\prime \prime },r^{\prime };E\right) =\int_{0}^{\infty
}dT\exp \left( \frac{i}{\hbar }ET\right) \widetilde{K}_{l}\left( r^{\prime
\prime },r^{\prime };T\right) ,  \label{3.7}
\end{equation}%
where the propagator $\widetilde{K}_{l}\left( r^{\prime \prime },r^{\prime
};T\right) $ is explicitly defined by%
\begin{eqnarray}
\widetilde{K}_{l}\left( r^{\prime \prime },r^{\prime };T\right)
&=&\lim_{N\rightarrow \infty }\left( \frac{\mu }{2i\pi \varepsilon \hbar }%
\right) ^{\frac{N+1}{2}}\dprod_{n=1}^{N}\left[ \int dr_{n}\right]  \notag \\
&&\times \exp \left\{ \frac{i}{\hbar }\underset{n=1}{\overset{N+1}{\sum }}%
\left[ \frac{\mu }{2\varepsilon }\left( \Delta r\right) _{n}^{2}-\varepsilon
V_{eff}(r_{n})\right] \right\} .  \label{3.8}
\end{eqnarray}

To construct the radial Green's function (\ref{3.7}), two cases occur
depending on whether $c_{h}$ is in the interval $]0.1[$ or in the interval $%
\left] -1,0\right[ $.

\section{Deformed Manning-Rosen potential}

If $0<c_{h}<1$, the Tietz-Wei potential turns into the Manning-Rosen
potential defined in terms of the $c_{h}-$deformed hyperbolic functions by%
\begin{equation}
V_{TW}\left( r\right) =V_{0}-V_{1}\coth _{c_{h}}\left[ \frac{b_{h}}{2}\left(
r-r_{e}\right) \right] +\frac{V_{2}}{\sinh _{c_{h}}^{2}\left[ \frac{b_{h}}{2}%
\left( r-r_{e}\right) \right] },  \label{3.9}
\end{equation}%
where $V_{0},V_{1}$ and $V_{2}$ are the real constants given by%
\begin{equation}
\left\{ 
\begin{array}{c}
V_{0}=\frac{D}{2}\left( 1+\frac{1}{c_{h}^{2}}\right) , \\ 
V_{1}=\frac{D}{2}\left( \frac{1}{c_{h}}+1\right) \left( \frac{1}{c_{h}}%
-1\right) , \\ 
V_{2}=\frac{D}{4}c_{h}\left( \frac{1}{c_{h}}-1\right) ^{2}.%
\end{array}%
\right.  \label{3.10}
\end{equation}%
The potential (\ref{3.9}) is obtained by using a $c_{h}-$deformation of the
usual hyperbolic functions denoted by

\begin{equation}
\left\{ 
\begin{array}{c}
\sinh _{c_{h}}x=\frac{e^{x}-c_{h}e^{-x}}{2},\text{ }\cosh _{c_{h}}x=\frac{%
e^{x}+c_{h}e^{-x}}{2},\text{ } \\ 
\tanh _{c_{h}}x=\frac{\sinh _{c_{h}}x}{\cosh _{c_{h}}x},\text{ }\coth
_{c_{h}}x=\frac{\cosh _{c_{h}}x}{\sinh _{c_{h}}x}.%
\end{array}%
\right.
\end{equation}%
These functions have been introduced for first time by Arai \cite{Arai},
with the real parameter\ $c_{h}>0$.

To evaluate the radial Green's function associated to the potential (\ref%
{3.9}), we have to distinguish two cases: $e^{-b_{h}r_{e}}\leq c_{h}<1$ et $%
0<$ $c_{h}<e^{-b_{h}r_{e}}.$

\subsection{First case: $e^{-b_{h}r_{e}}\leq c_{h}<1$}

The potential $V_{TW}\left( r\right) $ has a strong singularity at the point 
$r=r_{0}=r_{e}+\frac{1}{b_{h}}\ln c_{h}$. In this case, there are two
distinct regions, the first is defined by the interval $\left] 0,r_{0}\right[
$ and the second by the interval $\left] r_{0},+\infty \right[ $. We will be
interested in the calculation of the path integral for this potential only
in the interval $\left] r_{0},+\infty \right[ $, since, in the other
interval, the problem can not be solved analytically and moreover, its
solution does not have a physical interest worthy of note. In Figs. 1 and 2,
the potential (\ref{3.1}) is plotted in the interval $\left] r_{0},+\infty %
\right[ $ for different choices of the parameter $c_{h}$ of H$_{2}$(X$%
^{1}\Sigma _{g}^{+}$) and I$_{2}$(X(O$_{g}^{+}$)) molecules in the range $%
e^{-b_{h}r_{e}}\leq c_{h}<1$.

Now, to calculate the radial~Green's function for a state with an orbital
quantum number $l$, we adopt the expression \cite{Jia4}:%
\begin{equation}
\frac{1}{r^{2}}\approx C_{0}+\frac{B_{0}}{e^{b_{h}\left( r-r_{e}\right)
}-c_{h}}+\frac{A_{0}}{\left( e^{b_{h}\left( r-r_{e}\right) }-c_{h}\right)
^{2}}  \label{3.11}
\end{equation}%
as an approximation of the term $\frac{1}{r^{2}}$ contained in the
centrifugal potential term when $b_{h}\left( r-r_{e}\right) \ll 1$. Here $%
C_{0}=\frac{b_{h}^{2}}{12},$ $B_{0}$ and $A_{0}$ are two adjustable
parameters. If we take $C_{0}=0,B_{0}=\frac{A_{0}}{c_{h}},$ $%
A_{0}=b_{h}^{2}c_{h}^{2}$ and $c_{h}=1,$ the equation (\ref{3.11}) is
reduced to the approximation proposed by Greene and Aldrich \cite{Greene}.
Thus, after some simple calculation, we can write the expression of the
effective potential (\ref{3.4}) in the form: 
\begin{eqnarray}
V_{eff}\left( r\right) &\approx &V_{0}^{l}-V_{1}^{l}\coth _{c_{h}}\left[ 
\frac{b_{h}}{2}\left( r-r_{e}\right) \right] +\frac{V_{2}^{l}}{\sinh
_{c_{h}}^{2}\left[ \frac{b_{h}}{2}\left( r-r_{e}\right) \right] },  \notag \\
&&  \label{3.12}
\end{eqnarray}%
where the parameters $V_{0}^{l},V_{1}^{l}$ and $V_{2}^{l}$ are defined by%
\begin{equation}
\left\{ 
\begin{array}{c}
V_{0}^{l}=V_{0}+\frac{\hbar ^{2}l\left( l+1\right) }{2\mu }\left[ C_{0}+%
\frac{1}{2c_{h}}\left( \frac{A_{0}}{c_{h}}-B_{0}\right) \right] , \\ 
V_{1}^{l}=V_{1}+\frac{\hbar ^{2}l\left( l+1\right) }{4\mu c_{h}}\left( \frac{%
A_{0}}{c_{h}}-B_{0}\right) , \\ 
V_{2}^{l}=V_{2}+\frac{\hbar ^{2}l\left( l+1\right) A_{0}}{8\mu c_{h}}.%
\end{array}%
\right.  \label{3.13}
\end{equation}%
Then, by introducing the new variable $\xi =\frac{1}{2}\left[ b_{h}\left(
r-r_{e}\right) -\ln c_{h}\right] $ and putting $\varepsilon =\frac{b_{h}^{2}%
}{4}\varepsilon _{s}$, the Green's function$\ $(\ref{3.7}) is written%
\begin{equation}
G_{l}\left( r^{\prime \prime },r^{\prime };E\right) =\frac{2}{b_{h}}%
G_{MR}(\xi ^{\prime \prime },\xi ^{\prime };\widetilde{E}_{l}),  \label{3.14}
\end{equation}%
where%
\begin{equation}
G_{MR}(\xi ^{\prime \prime },\xi ^{\prime };\widetilde{E}_{l})=\int_{0}^{%
\infty }dS\exp \left( \frac{i}{\hbar }\frac{\widetilde{E}_{l}}{b_{h}^{2}}%
S\right) P_{l}\left( \xi ^{\prime \prime },\xi ^{\prime };S\right) ,
\label{3.15}
\end{equation}%
is an expression in which the kernel $P_{l}\left( \xi ^{\prime \prime },\xi
^{\prime };S\right) $ has the explicit form:%
\begin{eqnarray}
P_{l}\left( \xi ^{\prime \prime },\xi ^{\prime };S\right) &=&\frac{b_{h}}{2}%
\underset{N\rightarrow \infty }{\lim }\left( \frac{\mu }{2i\pi \hbar
\varepsilon _{s}}\right) ^{\frac{N+1}{2}}\overset{N}{\underset{n=1}{\dprod }}%
\left[ \int d\xi _{n}\right] \exp \left\{ \frac{i}{\hbar }\underset{n=1}{%
\overset{N+1}{\sum }}\left[ \frac{\mu }{2\varepsilon _{s}}\left( \Delta \xi
\right) _{n}^{2}\right. \right.  \notag \\
&&\left. \left. +\frac{4\varepsilon _{s}}{b_{h}^{2}}\left( V_{1}^{l}\coth
\xi _{n}-\frac{V_{2}^{l}}{c_{h}\sinh ^{2}\xi _{n}}\right) \right] \right\} ,
\notag \\
&&  \label{3.16}
\end{eqnarray}%
and%
\begin{equation}
\widetilde{E}_{l}=4\left( E-V_{0}^{l}\right) .  \label{3.17}
\end{equation}%
The expression (\ref{3.16}) is the one of the propagator for the standard
Manning-Rosen potential \cite{Manning} which has been discussed in the
literature by means of the path integral \cite{Grosche}. We can thus write
down the solution of (\ref{3.15}) immediately in a closed form as:%
\begin{eqnarray}
G_{MR}(\xi ^{\prime \prime },\xi ^{\prime };\widetilde{E}_{l}) &=&-\frac{%
i\mu }{\hbar }\frac{\Gamma \left( M_{1}-L_{E}\right) \Gamma \left(
L_{E}+M_{1}+1\right) }{\Gamma \left( M_{1}-M_{2}+1\right) \Gamma \left(
M_{1}+M_{2}+1\right) }  \notag \\
&&\times \left( \frac{2}{1+\coth \xi ^{\prime }}\frac{2}{1+\coth \xi
^{\prime \prime }}\right) ^{\frac{M_{1}+M_{2}+1}{2}}\left( \frac{\coth \xi
^{\prime }-1}{\coth \xi ^{\prime }+1}\frac{\coth \xi ^{\prime \prime }-1}{%
\coth \xi ^{\prime \prime }+1}\right) ^{\frac{M_{1}-M_{2}}{2}}  \notag \\
&&\times \text{ }_{2}F_{1}\left( M_{1}-L_{E},L_{E}+M_{1}+1,M_{1}-M_{2}+1;%
\frac{\coth \xi _{>}-1}{\coth \xi _{>}+1}\right)  \notag \\
&&\times \text{ }_{2}F_{1}\left( M_{1}-L_{E},L_{E}+M_{1}+1,M_{1}+M_{2}+1;%
\frac{2}{\coth \xi _{<}+1}\right) ,  \notag \\
&&  \label{3.18}
\end{eqnarray}%
where the parameters $L_{E},M_{1}$ and $M_{2}$ have in the present case the
following values:%
\begin{equation}
\left\{ 
\begin{array}{c}
L_{E}=-\frac{1}{2}+\frac{1}{\hbar b_{h}}\sqrt{2\mu \left(
V_{0}^{l}+V_{1}^{l}-E\right) }, \\ 
M_{1}=\delta _{l}+\frac{1}{\hbar b_{h}}\sqrt{2\mu \left(
V_{0}^{l}-V_{1}^{l}-E\right) }, \\ 
M_{2}=\delta _{l}-\frac{1}{\hbar b_{h}}\sqrt{2\mu \left(
V_{0}^{l}-V_{1}^{l}-E\right) },%
\end{array}%
\right.  \label{3.19}
\end{equation}%
with 
\begin{equation}
\delta _{l}=\sqrt{\frac{8\mu V_{2}}{\hbar ^{2}b_{h}^{2}c_{h}}+\frac{l(l+1)}{%
b_{h}^{2}c_{h}^{2}}A_{0}+\frac{1}{4}}.  \label{3.20}
\end{equation}%
So, going back to the old variable, the radial Green's function (\ref{3.14})
has for expression: 
\begin{eqnarray}
G_{l}\left( r^{\prime \prime },r^{\prime };E\right) &=&-\frac{2i\mu }{\hbar
b_{h}}\frac{\Gamma \left( M_{1}-L_{E}\right) \Gamma \left(
L_{E}+M_{1}+1\right) }{\Gamma \left( M_{1}-M_{2}+1\right) \Gamma \left(
M_{1}+M_{2}+1\right) }  \notag \\
&&\times \left[ \left( 1-c_{h}e^{-b_{h}\left( r^{\prime }-r_{e}\right)
}\right) \left( 1-c_{h}e^{-b_{h}\left( r^{\prime \prime }-r_{e}\right)
}\right) \right] ^{\frac{M_{1}+M_{2}+1}{2}}  \notag \\
&&\times \left( c_{h}^{2}e^{-b_{h}\left( r^{\prime }-r_{e}\right)
}e^{-b_{h}\left( r^{\prime \prime }-r_{e}\right) }\right) ^{\frac{M_{1}-M_{2}%
}{2}}  \notag \\
&&\times \text{ }_{2}F_{1}\left(
M_{1}-L_{E},L_{E}+M_{1}+1,M_{1}-M_{2}+1;c_{h}e^{-b_{h}\left(
r_{>}-r_{e}\right) }\right)  \notag \\
&&\times \text{ }_{2}F_{1}\left(
M_{1}-L_{E},L_{E}+M_{1}+1,M_{1}+M_{2}+1;1-c_{h}e^{-b_{h}\left(
r_{<}-r_{e}\right) }\right) .  \notag \\
&&  \label{3.21}
\end{eqnarray}

The energy spectrum for \ the bound states can be obtained from the poles of
the radial Green's function (\ref{3.21}). These poles are those of the gamma
function $\Gamma \left( M_{1}-L_{E}\right) $ that we find when its argument
is a negative integer or zero, i.e. when%
\begin{equation}
M_{1}-L_{E}=-n_{r},\text{ \ \ }n_{r}=0,1,2,3,....  \label{3.22}
\end{equation}%
Taking into account (\ref{3.19}), the eigenvalues of energy are then given by%
\begin{equation}
E_{n_{r},l}=V_{0}+\frac{\hbar ^{2}l(l+1)}{2\mu }\left[ C_{0}+\frac{1}{2c_{h}}%
\left( \frac{A_{0}}{c_{h}}-B_{0}\right) \right] -\frac{\hbar ^{2}b_{h}^{2}}{%
8\mu }\left( N_{r}^{2}+\frac{\lambda _{l}^{2}}{N_{r}^{2}}\right) ,
\label{3.23}
\end{equation}%
with the notation%
\begin{equation}
N_{r}=n_{r}+\delta _{l}+\frac{1}{2},  \label{3.24}
\end{equation}%
and%
\begin{equation}
\lambda _{l}=\frac{4\mu V_{1}}{\hbar ^{2}b_{h}^{2}}+\frac{l(l+1)}{%
b_{h}^{2}c_{h}}\left( \frac{A_{0}}{c_{h}}-B_{0}\right) .  \label{3.25}
\end{equation}

Now expressing the radial Green's function (\ref{3.21}) in the form of a
spectral expansion as follows:

\begin{equation}
G_{l}\left( r^{\prime \prime },r^{\prime };E\right) =i\hbar \overset{%
n_{r\max }}{\underset{n_{r}=0}{\sum }}\frac{\chi _{n_{r},l}\left( r^{\prime
\prime }\right) \chi _{n_{r},l}\left( r^{\prime }\right) }{E-E_{n_{r},l}},
\label{3.26}
\end{equation}%
we find for the wave functions $\chi _{n_{r},l}\left( r\right) $, suitably
normalized, the values:

\begin{eqnarray}
\chi _{n_{r},l}\left( r\right) &=&\mathcal{N}_{n_{r},l}\left(
1-c_{h}e^{-b_{h}(r-r_{e})}\right) ^{N_{r}-n_{r}}\left(
c_{h}e^{-b_{h}(r-r_{e})}\right) ^{\frac{1}{2}\left( \frac{\lambda _{l}}{N_{r}%
}-N_{r}\right) }  \notag \\
&&\times \text{ }_{2}F_{1}\left( -n_{r},N_{r}+\frac{\lambda _{l}}{N_{r}}%
-n_{r},\frac{\lambda _{l}}{N_{r}}-N_{r}+1;c_{h}e^{-b_{h}(r-r_{e})}\right) . 
\notag \\
&&
\end{eqnarray}%
where the normalization factor $\mathcal{N}_{n_{r},l}$ is%
\begin{eqnarray}
\mathcal{N}_{n_{r},l} &=&\left[ \frac{b_{h}}{2N_{r}}\left( \frac{\lambda _{l}%
}{N_{r}}+N_{r}\right) \left( \frac{\lambda _{l}}{N_{r}}-N_{r}\right) \frac{%
\Gamma \left( N_{r}-n_{r}+\frac{\lambda _{l}}{N_{r}}\right) \Gamma \left(
1-N_{r}+n_{r}+\frac{\lambda _{l}}{N_{r}}\right) }{n_{r}!\Gamma \left(
2N_{r}-n_{r}\right) }\right] ^{\frac{1}{2}}  \notag \\
&&\times \frac{1}{\Gamma \left( \frac{\lambda _{l}}{N_{r}}-N_{r}+1\right) }.
\end{eqnarray}%
Here note that $n_{r}=0,1,2,...,n_{r\max }<\sqrt{\lambda _{l}}-\delta _{l}-%
\frac{1}{2},$ with $n_{r\max }$ the maximum number of bound states.

In calculating energy values from the equation (\ref{3.23}) for some
molecules, it should be noted that the parameter $c_{h}$ must be greater
than or equal to the values in the table below and less than unity, unlike
what is claimed in the literature \cite{Falaye1}.

\begin{tabular}{|cccc|}
\hline
\multicolumn{4}{|c|}{Minimal values of the parameter $c_{h}$ for obtaining
the energy levels from Equation (\ref{3.23}).} \\ \hline
\multicolumn{1}{|c|}{molecule} & \ \ \ $b_{h}\left( \mathring{A}^{-1}\right) 
$ & \ \ $r_{e}\left( \mathring{A}\right) $ & \ \ \ \ $c_{h}$ \\ \hline
\multicolumn{1}{|c|}{HF(X$^{1}\Sigma ^{+}$)} & \ \ \ \ \ \ 1.942 07\ \ \ \ \
\  & \ 0.917 & 0.168 490 115 \\ \hline
\multicolumn{1}{|c|}{Cl$_{2}$(X$^{1}\Sigma _{g}^{+}$)\ } & 2.200 354 & 1.987
& 0.012 624 657 \\ \hline
\multicolumn{1}{|c|}{I$_{2}$(X(O$_{g}^{+}$))} & 2.123 43 & 2.666 & 0.003 478
812 \\ \hline
\multicolumn{1}{|c|}{H$_{2}$(X$^{1}\Sigma _{g}^{+}$) \ } & 1.618 90 & 0.741
& 0.301 313 237 \\ \hline
\multicolumn{1}{|c|}{O$_{2}$(X$^{3}\Sigma _{g}^{+}$)} & 2.591 03 & 1.207 & 
0.043 832 785 \\ \hline
\multicolumn{1}{|c|}{N$_{2}$(X$^{1}\Sigma _{g}^{+}$)} & 2.785 85 & 1.097 & 
0,047 071 975 \\ \hline
\multicolumn{1}{|c|}{CO(X$^{1}\Sigma ^{+}$)} & 2.204 81 & 1.128 & 0.083 156
934 \\ \hline
\multicolumn{1}{|c|}{NO(X$^{2}\Pi _{r}$)} & 2.715 59 & 1.151 & 0.043 908 643
\\ \hline
\multicolumn{1}{|c|}{O$_{2}^{+}$(X$^{2}\Pi _{g}^{+}$)} & 2.869 87 & 1.116 & 
0.040 649 248 \\ \hline
\multicolumn{1}{|c|}{NO$^{+}$(X$^{1}\Sigma ^{+}$)} & 2.733 49 & 1.063 & 
0.054 710 486 \\ \hline
\multicolumn{1}{|c|}{N$_{2}^{+}$(X$^{2}\Sigma _{g}^{+}$)} & 2.708 30 & 1.116
& 0.048 681 178 \\ \hline
\end{tabular}

\subsection{Second case: $0<c_{h}<e^{-b_{h}r_{e}}$~}

In this case, the potential (\ref{3.1}) is defined in the interval $%
%TCIMACRO{\U{211d} }%
%BeginExpansion
\mathbb{R}
%EndExpansion
^{+}.$ Figs. 3 and 4 represent the shape of the potential (\ref{3.1}) for
different values of $c_{h}$ in the range $0<c_{h}<e^{-b_{h}r_{e}}$~ for H$%
_{2}$(X$^{1}\Sigma _{g}^{+}$) and I$_{2}$(X(O$_{g}^{+}$)) molecules. The
transformation $r=r_{e}+\frac{1}{b_{h}}\left( 2\xi +\ln c_{h}\right) $
converts $r\in \left] 0,+\infty \right[ $ into $\xi \in \left] -\frac{1}{2}%
\left( b_{h}r_{e}+\ln c_{h}\right) ,+\infty \right[ .$ This means that the
kernel (\ref{3.16}) for $l=0$, is the propagator that describes the
evolution of a particle in the presence of a potential of the Manning-Rosen
type in the half-space $\xi >\xi _{0}=-\frac{1}{2}\left( b_{h}r_{e}+\ln
c_{h}\right) .$

Now, by performing the time transformation $\varepsilon =\frac{4\varepsilon
_{s}}{b_{h}^{2}}$, or $dT=\frac{4dS}{b_{h}^{2}}$, we can rewrite (\ref{3.7}%
), for $l=0$, as

\begin{equation}
G_{0}(r^{\prime \prime },r^{\prime };E)=\frac{2}{b_{h}}\widetilde{G}%
_{MR}\left( \xi ^{\prime },\xi ^{\prime \prime };\tilde{E}_{0}\right) ,
\label{3.28}
\end{equation}%
with\ 
\begin{equation}
\widetilde{G}_{MR}\left( \xi ^{\prime },\xi ^{\prime \prime };\tilde{E}%
_{0}\right) =\underset{0}{\overset{\infty }{\dint }}dS\exp \left( \frac{i}{%
\hbar }\frac{\tilde{E}_{0}}{b_{h}^{2}}S\right) P_{0}(\xi ^{\prime \prime
},\xi ^{\prime };S),  \label{3.29}
\end{equation}%
where

\begin{equation}
\tilde{E}_{0}=4\left( E-V_{0}\right) ,  \label{3.30}
\end{equation}%
and%
\begin{eqnarray}
P_{0}\left( \xi ^{\prime \prime },\xi ^{\prime };S\right) &=&\frac{b_{h}}{2}%
\underset{N\rightarrow \infty }{\lim }\left( \frac{\mu }{2i\pi \hbar
\varepsilon _{s}}\right) ^{\frac{N+1}{2}}\overset{N}{\underset{n=1}{\dprod }}%
\left[ \int d\xi _{n}\right] \exp \left\{ \frac{i}{\hbar }\sum_{n=1}^{N+1}%
\left[ \frac{\mu }{2\varepsilon _{s}}\left( \Delta \xi \right)
_{n}^{2}\right. \right.  \notag \\
&&+\left. \left. \frac{4\varepsilon _{s}}{b_{h}^{2}}\left( V_{1}\coth \xi
_{n}-\frac{V_{2}}{c_{h}\sinh ^{2}\xi _{n}}\right) \right] \right\} .\text{ }
\notag \\
&&  \label{3.31}
\end{eqnarray}%
Note that the kernel (\ref{3.31}) represents the propagator relative to the
Manning-Rosen potential defined in the half-space $\xi >\xi _{0}$ \cite%
{Khodja1}. The details concerning the evaluation of the Green's function (%
\ref{3.29}) in terms of the Green's function in the interval $%
%TCIMACRO{\U{211d} }%
%BeginExpansion
\mathbb{R}
%EndExpansion
^{+}$ through the perturbation expansion method discussed in the literature 
\cite{Grosche2,Benamira1,Benamira2}\ are similar to those presented in our
previous work on the vector and scalar deformed radial
Rosen-Morse-potentials \cite{Khodja2}. Hence we obtain

\begin{equation}
\widetilde{G}_{MR}\left( \xi ^{\prime \prime },\xi ^{\prime };\tilde{E}%
_{0}\right) =G_{MR}^{0}(\xi ^{\prime \prime },\xi ^{\prime };\tilde{E}_{0})-%
\frac{G_{MR}^{0}(\xi ^{\prime \prime },\xi _{0};\tilde{E}_{0})G_{MR}^{0}(\xi
_{0},\xi ^{\prime };\tilde{E}_{0})}{G_{MR}^{0}(\xi _{0},\xi _{0};\tilde{E}%
_{0})},  \label{3.32}
\end{equation}%
where $G_{MR}^{0}(\xi ^{\prime \prime },\xi ^{\prime };\tilde{E}_{0})$ is
the Green's function (\ref{3.18}) associated with the standard Manning-Rosen
potential, for the $s-$waves $(l=0)$ and $0<c_{h}<e^{-b_{h}r_{e}}.$

The poles of the Green's function (\ref{3.32}) yield the energy eigenvalues $%
E_{n_{r}}$. These poles are determined by the equation $G_{MR}^{0}(\xi
_{0},\xi _{0};\tilde{E}_{0})=0$ which immediatey leads to the following
transcendental equation

\begin{equation}
_{2}F_{1}\left(
M_{1}-L_{E_{n_{r}}},L_{E_{n_{r}}}+M_{1}+1,M_{1}-M_{2}+1;c_{h}e^{b_{h}r_{e}}%
\right) =0,  \label{3.33}
\end{equation}%
where the quantities $L_{E_{n_{r}}},M_{1}$ and $M_{2}$ are defined by

\begin{equation}
\left\{ 
\begin{array}{c}
L_{E_{n_{r}}}=-\frac{1}{2}+\frac{1}{\hbar b_{h}}\sqrt{2\mu \left(
V_{0}+V_{1}-E_{n_{r}}\right) }, \\ 
M_{1}=\delta _{0}+\frac{1}{\hbar b_{h}}\sqrt{2\mu \left(
V_{0}-V_{1}-E_{n_{r}}\right) }, \\ 
M_{2}=\delta _{0}-\frac{1}{\hbar b_{h}}\sqrt{2\mu \left(
V_{0}-V_{1}-E_{n_{r}}\right) }.%
\end{array}%
\right.  \label{3.34}
\end{equation}%
with

\begin{equation}
\delta _{0}=\sqrt{\frac{1}{4}+\frac{8\mu V_{2}}{\hbar ^{2}b_{h}^{2}c_{h}}}.
\label{3.35}
\end{equation}

The energy levels $E_{n_{r}}$ of \ the molecule are given by the solutions
of the transcendental equation (\ref{3.33}), which are to be found from
numerical method and the corresponding wave functions will have the form%
\begin{eqnarray}
\mathcal{\chi }_{n_{r}}^{0<c_{h}<e^{-b_{h}r_{e}}}(r) &=&\mathcal{N}\left(
1-c_{h}e^{-b_{h}\left( r-r_{e}\right) }\right) ^{\delta _{0}+\frac{1}{2}}%
\text{ }\left( c_{h}e^{-b_{h}\left( r-r_{e}\right) }\right) ^{\frac{1}{\hbar
b_{h}}\sqrt{2\mu \left( V_{0}+V_{1}-E_{n_{r}}\right) }}  \notag \\
&&\times _{2}F_{1}\left(
M_{1}-L_{E_{n_{r}}},L_{E_{n_{r}}}+M_{1}+1,M_{1}-M_{2}+1;c_{h}e^{-b_{h}\left(
r-r_{e}\right) }\right) ,  \notag \\
&&  \label{3.36}
\end{eqnarray}%
where $\mathcal{N}$ is a constant factor. Note that these wave functions
satisfy the boundary conditions:

\begin{equation}
\mathcal{\chi }_{n_{r}}^{0<c_{h}<e^{-b_{h}r_{e}}}(r)\underset{r\rightarrow 0}%
{\longrightarrow 0},  \label{3.37}
\end{equation}%
and%
\begin{equation}
\mathcal{\chi }_{n_{r}}^{0<c_{h}<e^{-b_{h}r_{e}}}(r)\underset{r\rightarrow
\infty }{\longrightarrow 0}.  \label{3.38}
\end{equation}

\section{Deformed Rosen-Morse potential}

When $-1<c_{h}<0$, the Tietz-Wei potential matches the form of the
Rosen-Morse potential expressed in terms of the $c_{h}-$deformed hyperbolic
functions by 
\begin{equation}
V_{TW}(r)=U_{0}+U_{1}\tanh _{\left\vert c_{h}\right\vert }\left[ \frac{b_{h}%
}{2}\left( r-r_{e}\right) \right] -\frac{U_{2}}{\cosh _{\left\vert
c_{h}\right\vert }^{2}\left[ \frac{b_{h}}{2}\left( r-r_{e}\right) \right] },
\label{3.39}
\end{equation}%
where the constants $U_{0},U_{1}$ and $U_{2}$ are defined by%
\begin{equation}
\left\{ 
\begin{array}{c}
U_{0}=\frac{D}{2}\left( 1+\frac{1}{\left\vert c_{h}\right\vert ^{2}}\right) ,
\\ 
U_{1}=\frac{D}{2}\left( 1-\frac{1}{\left\vert c_{h}\right\vert }\right)
\left( 1+\frac{1}{\left\vert c_{h}\right\vert }\right) , \\ 
U_{2}=\frac{D}{4}\left\vert c_{h}\right\vert \left( \frac{1}{\left\vert
c_{h}\right\vert }+1\right) ^{2}.%
\end{array}%
\right.  \label{3.40}
\end{equation}%
The potential (\ref{3.1}) is plotted in Figs. 5 and 6 for different values
of $c_{h}$ in the range $-1<c_{h}<0$ for H$_{2}$(X$^{1}\Sigma _{g}^{+}$) and
I$_{2}$(X(O$_{g}^{+}$)) molecules.

In this case, the radial Green's function associated with the waves $s(l=0)$
is written%
\begin{equation}
G\left( r^{\prime \prime },r^{\prime };\mathcal{E}_{0}\right)
=\int_{0}^{\infty }dT\exp \left( \frac{i}{\hbar }\mathcal{E}_{0}T\right)
K_{0}\left( r^{\prime \prime },r^{\prime };T\right)  \label{3.41}
\end{equation}%
with%
\begin{equation}
\mathcal{E}_{0}=E-U_{0},
\end{equation}%
and%
\begin{eqnarray}
K_{0}\left( r^{\prime \prime },r^{\prime };T\right) &=&\int \mathcal{D}%
r\left( t\right) \exp \left\{ \frac{i}{\hbar }\int_{0}^{T}\left[ \frac{\mu }{%
2}\overset{.}{r}^{2}-\left( U_{1}\tanh _{\left\vert c_{h}\right\vert }\left[ 
\frac{b_{h}}{2}\left( r-r_{e}\right) \right] \right. \right. \right.  \notag
\\
&&\left. \left. \left. -\frac{U_{2}}{\cosh _{\left\vert c_{h}\right\vert
}^{2}\left[ \frac{b_{h}}{2}\left( r-r_{e}\right) \right] }\right) \right]
dt\right\} .  \label{3.42}
\end{eqnarray}

Now, by applying the spatial transformation $r\in 
%TCIMACRO{\U{211d} }%
%BeginExpansion
\mathbb{R}
%EndExpansion
^{+}\rightarrow x\in \left] x_{0},+\infty \right[ $ defined by%
\begin{equation}
x=\frac{b_{h}}{2}\left( r-r_{e}\right) -\frac{1}{2}\ln \left\vert
c_{h}\right\vert ,  \label{3.43}
\end{equation}%
accompanied by the time transformation 
\begin{equation}
\varepsilon =\frac{4}{b_{h}^{2}}\varepsilon _{s}\text{ ou \ }T=\frac{4}{%
b_{h}^{2}}S,  \label{3.44}
\end{equation}%
we can thus rewrite the function of Green (\ref{3.41}) in the following form:%
\begin{equation}
G\left( r^{\prime \prime },r^{\prime };\mathcal{E}_{0}\right) =\frac{2}{b_{h}%
}\widetilde{G}\left( x^{\prime \prime },x^{\prime };\mathcal{E}_{0}\right) ,
\label{3.45}
\end{equation}%
with%
\begin{equation}
\widetilde{G}\left( x^{\prime \prime },x^{\prime };\mathcal{E}_{0}\right)
=\int_{0}^{\infty }dS\exp \left( \frac{4i}{\hbar b_{h}^{2}}\mathcal{E}%
_{0}S\right) K_{0}\left( x^{\prime \prime },x^{\prime };\mathcal{E}%
_{0}\right) ,  \label{3.46}
\end{equation}%
where 
\begin{equation}
K_{0}\left( x^{\prime \prime },x^{\prime };\mathcal{E}_{0}\right) =\int 
\mathcal{D}x(s)\exp \left\{ \frac{i}{\hbar }\int_{0}^{S}\left[ \frac{\mu }{2}%
\overset{.}{x}^{2}-\frac{4}{b_{h}^{2}}\left( U_{1}\tanh x-\frac{U_{2}}{%
\left\vert c_{h}\right\vert \cosh ^{2}x}\right) \right] ds\right\} ;\text{ \ 
}x>x_{0}  \label{3.47}
\end{equation}%
is the propagator associated with a particle that moves in the half-space $%
x>x_{0}=-\frac{1}{2}\left( b_{h}r_{e}+\ln \left\vert c_{h}\right\vert
\right) $, under the action of the standard Rosen-Morse potential \cite%
{Rosen}. Since the propagator (\ref{3.47}) is defined in the half-space $%
x>x_{0}$, we have to construct the corresponding Green's function (\ref{3.46}%
) in terms of the Green's function for that potential in all the space $%
%TCIMACRO{\U{211d} }%
%BeginExpansion
\mathbb{R}
%EndExpansion
$ by proceeding as in the previous case. All calculations done, we obtain 
\begin{equation}
\widetilde{G}\left( x^{\prime \prime },x^{\prime };\mathcal{E}_{0}\right)
=G_{RM}\left( x^{\prime \prime },x^{\prime };\mathcal{E}_{0}\right) -\frac{%
G_{RM}\left( x^{\prime \prime },x_{0};\mathcal{E}_{0}\right) G_{RM}\left(
x_{0},x^{\prime };\mathcal{E}_{0}\right) }{G_{RM}\left( x_{0},x_{0};\mathcal{%
E}_{0}\right) },  \label{3.48}
\end{equation}%
where $G_{RM}\left( x^{\prime \prime },x^{\prime };\mathcal{E}_{0}\right) $
is the Green's function relative to the standard Rosen-Morse potential \cite%
{Kleinert} given by

\begin{eqnarray}
G_{RM}\left( x^{\prime \prime },x^{\prime };\mathcal{E}_{0}\right) &=&-\frac{%
i\mu }{\hbar }\frac{\Gamma \left( M_{1}-L_{\mathcal{E}_{0}}\right) \Gamma
\left( L_{\mathcal{E}_{0}}+M_{1}+1\right) }{\Gamma \left(
M_{1}+M_{2}+1\right) \Gamma \left( M_{1}-M_{2}+1\right) }  \notag \\
&&\times \left( \frac{1-\tanh x^{\prime }}{2}.\frac{1-\tanh x^{\prime \prime
}}{2}\right) ^{\frac{M_{1}+M_{2}}{2}}  \notag \\
&&\times \left( \frac{1+\tanh x^{\prime }}{2}.\frac{1+\tanh x^{\prime \prime
}}{2}\right) ^{\frac{M_{1}-M_{2}}{2}}  \notag \\
&&\times \text{ }_{2}F_{1}\left( M_{1}-L_{\mathcal{E}_{0}},L_{\mathcal{E}%
_{0}}+M_{1}+1,M_{1}-M_{2}+1;\frac{1+\tanh x_{<}}{2}\right)  \notag \\
&&\times \text{ }_{2}F_{1}\left( M_{1}-L_{\mathcal{E}_{0}},L_{\mathcal{E}%
_{0}}+M_{1}+1,M_{1}+M_{2}+1;\frac{1-\tanh x_{>}}{2}\right) .  \notag \\
&&  \label{3.49}
\end{eqnarray}%
Here we have used the following abbreviations%
\begin{equation}
\left\{ 
\begin{array}{c}
L_{\mathcal{E}_{0}}=-\frac{1}{2}+\sqrt{\frac{1}{4}+\frac{8\mu U_{2}}{\hbar
^{2}b_{h}^{2}\left\vert c_{h}\right\vert }}, \\ 
M_{1}=\frac{1}{\hbar b_{h}}\sqrt{2\mu \left( D-E\right) }+\frac{1}{\hbar
b_{h}}\sqrt{2\mu \left( \frac{D}{\left\vert c_{h}\right\vert ^{2}}-E\right) }%
, \\ 
M_{2}=\frac{1}{\hbar b_{h}}\sqrt{2\mu \left( D-E\right) }-\frac{1}{\hbar
b_{h}}\sqrt{2\mu \left( \frac{D}{\left\vert c_{h}\right\vert ^{2}}-E\right) }%
.%
\end{array}%
\right.  \label{3.50}
\end{equation}

The bound states with the energy $E_{n_{r}}$ are determined from the poles
of the Green's function $G_{RM}\left( x_{0},x_{0};\mathcal{E}_{0}\right) $
by the transcendental equation 
\begin{equation}
_{2}F_{1}\left( M_{1}-L_{\mathcal{E}_{0}},L_{\mathcal{E}%
_{0}}+M_{1}+1,M_{1}+M_{2}+1;\frac{\left\vert c_{h}\right\vert }{%
e^{-b_{h}r_{e}}+\left\vert c_{h}\right\vert }\right) =0\text{ \ \ \ for }%
E=E_{n_{r}},  \label{3.51}
\end{equation}%
which can also be solved numerically and the corresponding wave functions
have the form:%
\begin{eqnarray}
\chi _{n_{r}}\left( r\right) &=&\mathcal{N}\left( \frac{\left\vert
c_{h}\right\vert }{\left\vert c_{h}\right\vert +e^{b_{h}\left(
r-r_{e}\right) }}\right) ^{\frac{1}{\hbar b_{h}}\sqrt{2\mu \left(
D-E_{n_{r}}\right) }}\left( \frac{1}{1+\left\vert c_{h}\right\vert
e^{-b_{h}\left( r-r_{e}\right) }}\right) ^{\frac{1}{\hbar b_{h}}\sqrt{2\mu
\left( \frac{D}{\left\vert c_{h}\right\vert ^{2}}-E_{n_{r}}\right) }}  \notag
\\
&&\times \text{ }_{2}F_{1}\left( M_{1}-L_{\mathcal{E}_{0}},L_{\mathcal{E}%
_{0}}+M_{1}+1,M_{1}+M_{2}+1;\frac{\left\vert c_{h}\right\vert }{%
e^{b_{h}\left( r-r_{e}\right) }+\left\vert c_{h}\right\vert }\right) ,
\label{3.52}
\end{eqnarray}%
where $\mathcal{N}$ is a constant factor.

\section{Morse potential}

By making $\left\vert c_{h}\right\vert =0$ in the expression (\ref{3.1}), we
obtain the radial Morse potential \cite{Morse}

\begin{equation}
V_{M}(r)=D\left( 1-e^{-\beta \left( r-r_{e}\right) }\right) ^{2}.
\label{3.53}
\end{equation}%
In this case, we can show from the equations (\ref{3.50}) that

\begin{equation}
\left\{ 
\begin{array}{c}
L_{\mathcal{E}_{0}}\underset{\left\vert c_{h}\right\vert \rightarrow 0}{%
\simeq }-\frac{1}{2}+\left( 1+\frac{1}{\left\vert c_{h}\right\vert }\right) 
\frac{\sqrt{2\mu D}}{\hbar \beta }, \\ 
M_{1}\underset{\left\vert c_{h}\right\vert \rightarrow 0}{\simeq }\frac{1}{%
\hbar \beta }\sqrt{2\mu \left( D-E\right) }+\frac{\sqrt{2\mu D}}{\hbar \beta
\left\vert c_{h}\right\vert }, \\ 
M_{2}\underset{\left\vert c_{h}\right\vert \rightarrow 0}{\simeq }\frac{1}{%
\hbar \beta }\sqrt{2\mu \left( D-E\right) }-\frac{\sqrt{2\mu D}}{\hbar \beta
\left\vert c_{h}\right\vert }.%
\end{array}%
\right.  \label{3.54}
\end{equation}

On the other hand, taking into account the property of the hypergeometric
function \cite{Landau}, it is easy to see that 
\begin{eqnarray}
&&\lim_{\left\vert c_{h}\right\vert \rightarrow 0}\text{ }_{2}F_{1}\left(
M_{1}-L_{\mathcal{E}_{0}},L_{\mathcal{E}_{0}}+M_{1}+1,M_{1}+M_{2}+1;\frac{%
\left\vert c_{h}\right\vert }{e^{-b_{h}r_{e}}+\left\vert c_{h}\right\vert }%
\right)  \notag \\
&=&\text{ }_{1}F_{1}\left( \frac{1}{2}-\frac{\sqrt{2\mu D}}{\hbar \beta }+%
\frac{1}{\hbar \beta }\sqrt{2\mu \left( D-E\right) },\frac{2}{\hbar \beta }%
\sqrt{2\mu \left( D-E\right) }+1;2\frac{\sqrt{2\mu D}}{\hbar \beta }e^{\beta
r_{e}}\right) =0.  \notag \\
&&  \label{3.55}
\end{eqnarray}

By considerations similar to those used in Refs. \cite{Flugge, Guechi}, it
is easy to show that%
\begin{equation}
\frac{1}{2}-\frac{\sqrt{2\mu D}}{\hbar \beta }+\frac{1}{\hbar \beta }\sqrt{%
2\mu \left( D-E\right) }=-n_{r},\text{ \ \ \ \ \ \ \ }n_{r}=0,1,2,....
\label{3.56}
\end{equation}%
From where we find the well-known energy levels associated with the radial
potential of Morse:

\begin{equation}
E_{n_{r}}=-\frac{\hbar ^{2}\beta ^{2}}{2\mu }\left[ \left( n_{r}+\frac{1}{2}%
\right) ^{2}-2\left( n_{r}+\frac{1}{2}\right) \frac{\sqrt{2\mu D}}{\hbar
\beta }\right] ,  \label{3.57}
\end{equation}%
with $n_{r\text{ }\max }=\left\{ \frac{\sqrt{2\mu D}}{2\hbar \beta }-\frac{1%
}{2}\right\} $, and from the expression (\ref{3.52}), passing to the limit
where $\left\vert c_{h}\right\vert \rightarrow $ $0$, we will have for the
corresponding wave functions:

\begin{eqnarray}
\chi _{n_{r}}^{\left\vert c_{h}\right\vert =0}\left( r\right) &=&\mathcal{C}%
\exp \left[ -\frac{\sqrt{2\mu D}}{\hbar \beta }e^{-\beta \left(
r-r_{e}\right) }\right] \left( e^{-\beta \left( r-r_{e}\right) }\right) ^{%
\frac{\sqrt{2\mu D}}{\hbar \beta }-n_{r}-\frac{1}{2}}  \notag \\
&&\times \text{ }_{1}F_{1}\left( -n_{r},-2n_{r}+2\frac{\sqrt{2\mu D}}{\hbar
\beta };\frac{2\sqrt{2\mu D}}{\hbar \beta }e^{-\beta \left( r-r_{e}\right)
}\right) ,  \label{3.58}
\end{eqnarray}%
where $\mathcal{C}$ is the normalization constant.

\section{Conclusion}

In this paper, we have seen that the problem of the Tietz-Wei potential is
completely solved in the context of Feynman's path integrals, contrary to
the naive attempts of its treatment by various techniques \cite%
{Hamzavi1,Hamzavi2,Falaye1,Falaye2,Mikulski,Sameer2,Zarrinkamar} leading to
partially acceptable results. It should be noted that a suitable study of
this potential requires distinguishing three cases representing the standard
Manning-Rosen potential for $e^{-b_{h}r_{e}}\leq c_{h}<1$, the defined
Manning-Rosen potential in a half-space when $0<c_{h}<e^{-b_{h}r_{e}}$ and
the Rosen-Morse potential for $-1<c_{h}<0$. Each case requires special
handling. This proves that the optimization parameter $c_{h\text{ }}$ is an
important parameter in the analysis of this problem. In particular, the
analytical energy spectrum and the wave functions of the states $l$ of
certain diatomic molecules are obtained approximately for values {}{}of $c_{h%
\text{ }}$ greater than or equal to those contained in the table above. It
should also be pointed out that the\ Pekeris approximation with dependent
constants of $c_{h\text{ }}$ used in the works cited above applies very
restrictively to diatomic molecules. In the last two cases, the perturbation
expansion technique we did not repeat here is used to construct the Green's
function in a closed form. In each case, the poles of the Green's function
yield a transcendental equation involving the hypergeometric function that
must be solved numerically to determine the bound state energy levels $%
E_{n_{r}}$.

\end{document}